# Watertightization of Trimmed Surfaces at Intersection Boundary

Hua Li, Lu Zhang, Ruoxi Guo, Zushang Xiao, and Rui Guo
Cocasoft.com, 100190 Beijing, China
{huali@cocasoft.cn}

**Abstract:**

This paper introduces a watertight technique to deal with the boundary representation of surface-surface intersection in CAD.

Surfaces play an important role in today's geometric design. The mathematical model of non-uniform rational B-spline surfaces (NURBS) is the mainstream and ISO standard. In the situation of surface-surface intersection, things are a little complicated, for some parts of surfaces may be cut-off, so called trimmed surfaces occur, which is the central topic in the past decades in CAD community of both academia and industry. The main problem is that the parametric domain of the trimmed surface generally is not the standard square or rectangle, and rather, typically, bounded by curves, based on point inverse of the intersection points and interpolated. The existence of gaps or overlaps at the intersection boundary makes hard the preprocessing of CAE and other downstream applications. The NURBS are in this case hard to keep a closed form. In common, a special data structure of intersection curves must be affiliated to support downstream applications, while the data structure of the whole CAD system is not unified, and the calculation is not efficient.

In terms of Bezier surface, a special case of NURBS, this paper designs a reparameterization or normalization to transform the trimmed surface into a group of Bezier surface patches in standard parametric domain [0,1]X[0,1]. And then the boundary curve of normalized Bezier surface patch can be replaced by the intersection curve to realize watertight along the boundary. In this way, the trimmed surface is wiped out, the "gap" between CAD and CAE is closed.

# 1. Introduction

Geometric models are widely used in CAD and other related fields to represent the shape of products and/or objects. These geometric models provide precise descriptions of three-dimensional shapes in a way that computers can understand. Most geometric models are designed using parametric surfaces and stored in specific data structures, which greatly improves the efficiency of product model design, generation, simulation, modification and optimization. Most of the current mainstream geometric models use spline surface models, which represent the three-dimensional shape of products or objects with a series of regular surfaces and/or irregular surfaces. The non-uniform rational B-splines (NURBS) method in the form of tensor product is the mainstream technology for surface representation in CAD and is already universal and ISO standard, which plays an important and indispensable role in the design, analysis and manufacturing of industrial products. Bezier surface is a special expression form of NURBS. All NURBS surfaces can be represented by Bezier surface patches. Bezier surface patches are defined in the standard parametric domain [0,1]X[0,1], which is convenient for calculation, storage and transformation. The de Casteljau algorithm provides a tool for extracting partial surface patches along rectangular subregions surrounded by arbitrary isoparametric lines for Bezier surface patches. Each of these subregions can be transformed back into standard domains through linear parametric transformation [Böhm et al.,1984]. For the convenience of writing, the following explanation assumes that all spline surfaces and spline curves are converted into Bezier form.

In actual product design, complex surface shapes are often involved, they cannot be directly represented by a single spline surface. Instead, the trimmed surface is obtained by manipulating the intersection curves between the surfaces and then trimming and combining them. The processing of intersection curves between surfaces will affect the quality and accuracy of product design. If the intersection curves are not handled well, there will be gaps or overlaps in the surface boundaries of the product, causing the definition of geometric entities to not be closed on the boundaries, affecting downstream applications, like analysis and processing manufacturing. Therefore, how to deal with the intersection curves between surfaces has always been one of the core issues in CAD. [Farin,2002]

According to algebraic theory, the algebraic degree of a Bezier surface with parametric degree MXN is 2MN. For the common 3X3 parametric Bezier surface, its algebraic degree is 18. When two such surfaces intersect, the algebraic degree of the intersection curve is as high as 324, which results in no analytical solution in algebra and cannot be expressed accurately. Therefore, the intersection curve between two tensor product parametric surfaces can generally only be approximately represented within a certain accuracy by a low-order spline curve. The error or bias causes the intersection curve between two intersection surfaces to generally not strictly fit on either of the two intersection surfaces, except for the intersection point. That is to say, generally there is a gap or overlap between the intersection curve of the low-order approximation and the two intersection surfaces, so the definition of the product geometric model is incomplete and not closed. This problem is called "watertight" problem. This is one of the main sources of so-called "dirty geometry" data that results in imperfect, impure geometric designs. [Kasik et al., 2005], [Piegl,2005], [Sederberg et al., 2008], [Cottrell et al.,2009]

Therefore, various treatment methods have been derived. At present, the common method to deal with the "watertight" problem in CAD is to perform meshing at the boundaries and manually stitch the boundary meshes to achieve watertightness. However, there is no unified standard for manual stitching of seams. The stitching results vary from person to person and are approximate. For product designs with high precision requirements, premature meshing will also introduce unnecessary errors into subsequent simulation analysis. In addition, manually stitching the boundary mesh is tedious, brings additional design burden to the design engineer, requires a lot of time and attention, and affects the design efficiency of the product. Isogeometric analysis (IGA) aims at solving the connection of CAD and CAE with high order continuity functions and is now a hot research topic. [Hughes et al.,2005], [Cottrell et al.,2009], [Marussig and Hughes,2018]

A new idea is to use the T-spline method. While it introduces new mathematical model and its theory is not mature enough for convenient engineering use and is not an ISO standard. The difficulty and complexity of software development and maintenance are greatly increased, making it inconvenient to apply. It is not currently adopted by mainstream CAD systems yet. [Sederberg et al., 2003, 2004]

Another method is to seek some kind of transformation on the boundary, such as [Urick et al., 2019], which uses a linear transformation of separating multiple isoparametric lines to simply transform the trimmed boundary curve on an interval into a straight line of isoparametric lines. However, due to the lack of constraints defined by the surface, its impact on the shape of the original surface cannot be controlled accurately, and this problem has not been effectively addressed. [Urick, 2016], [Urick et al. 2019, 2020]

In terms of Bezier surface, this paper designs a reparameterization or normalization to transform the trimmed surface into a group of Bezier surface patches in standard parametric domain [0,1]X[0,1]. And then the boundary curve of normalized Bezier surface patch is replaced by the intersection curve to realize watertight. In this way, the trimmed surface is wiped out, the "gap" between CAD and CAE is closed.

## 2.Foundamentals, trimmed surfaces and its parametric domain

A general MXN Bezier surface is:

$$S(u,v) = \sum_{i=0}^{m}\sum_{j=0}^{n} B_{i,m}(u)B_{j,n}(v)R_{i,j} \tag{1}$$

where: $R_{i,j} \in R^3$ control points; $B_{k,l}(x) = \binom{l}{k}(1-x)^{l-k}x^k$, $\binom{l}{k} = \frac{l!}{(l-k)!k!}$, is the kth l degree Bernstein basis function of variable $x \in [0,1]$.

Assume that two Bezier surfaces S1(u,v), S2(s,t) intersect on a space curve, as shown in Figure 1. Since the precise mathematical representation of the intersection curve cannot be obtained, generally only a series of intersection points in the three-dimensional space could be given, say $P_i$, i=1,2,...,K. Through these intersection points, an interpolated spline representation of the intersection curve C(w)={x(w),y(w),z(w)} can be generated. Since the intersection points are on the two surfaces, the corresponding surface parameters can be found in reverse, $(u_j,v_j)$ and $(s_j, t_j)$, j=1,2,...,K. At the same time, the parametric splines on the two-dimensional parametric spaces of two surfaces can be generated respectively, C1(α) ={u(α),v(α)} and C2(β)={s(β),t(β)}, they are called domain curves.

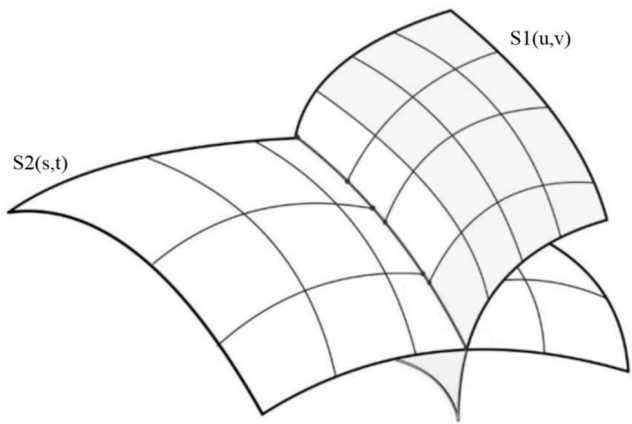

Figure 1 Two intersection surfaces and their intersection curve with defining points $P_j$.

By bringing the domain curves C1 and C2 into surfaces S1 and S2 respectively, we can obtain the curves SC1 and SC2 mapped to the two surfaces in the three-dimensional space, which are completely fitted on their respective surfaces, they are curves on the surfaces.

Note that there exist three curves on the boundary, C, SC1, and SC2, they are independent with each other and do not coincide exactly with each other. The gaps or overlaps there are the reason of "watertight problem".

## 3. Domain transformation and normalization

For the trimmed surface, the domain curve goes through the parametric domain and divides it into two parts, one is left, another will be cut-off, see Figure 2, where suppose the upright part be cut-off, point B, D, and G be corresponding points or projective points of intersection points in the domain.

On the left area of domain, the trimmed Bezier surface can be further segmented into a series of small surface patches according to the intersection points parameters at the boundary. These small patches can be classified into two categories, one is non-boundary patch, their parametric domains are defined by isoparametric lines; the other is boundary patch, their parametric domain contains some part of C1(α) or C2(β), which is composed of a trapezoid with curved sides. As shown in Figure 2, ABCD and DEFG are curved trapezoids with curved boundaries.

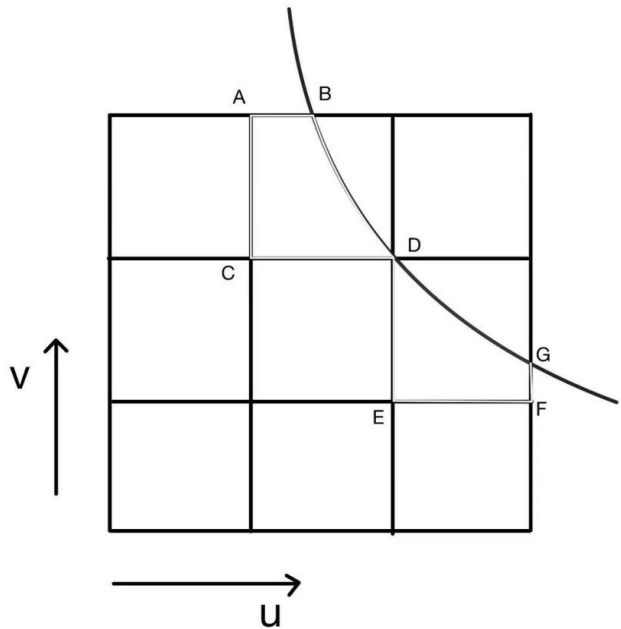

Figure 2 Parametric domain segmentation of trimmed surface.

Note that the domain curves C1(α) and C2(β) may be multi-valued functions. For example, if the intersection curve on the surface S1(u,v) is approximately a circle, then for C1 in the parametric domain, for each value of u (or v), there are two corresponding v (or u). The parametric domain may need to be appropriately decomposed further so that it contains only single-valued functions in the sub-region. This is always possible .

The Bezier surface shown in formula (1) is defined in the standard parametric domain [0,1]X[0,1], which is a canonical form. For the sub-domains of non-boundary patches, de Casteljau algorithm can be used to transform them into standard domain.

For the boundary patches, a parametric transformation is designed to restore their curved trapezoid domains into the standard one, and make the irregular domain to regular domain, this is called reparameterization or normalization.

The analysis shows that after segmentation above, there are eight types for the curved trapezoid domain as shown in Figure 3(a)-Figure 3(h), where there has one and only one point of domain curve passing through one vertex of the rectangular area, they are equivalent under a rotational symmetry transformation. The curved trapezoid ABCD and DEFG in Figure 2 corresponds to Figure 3(a) and Figure 3(c) respectively.

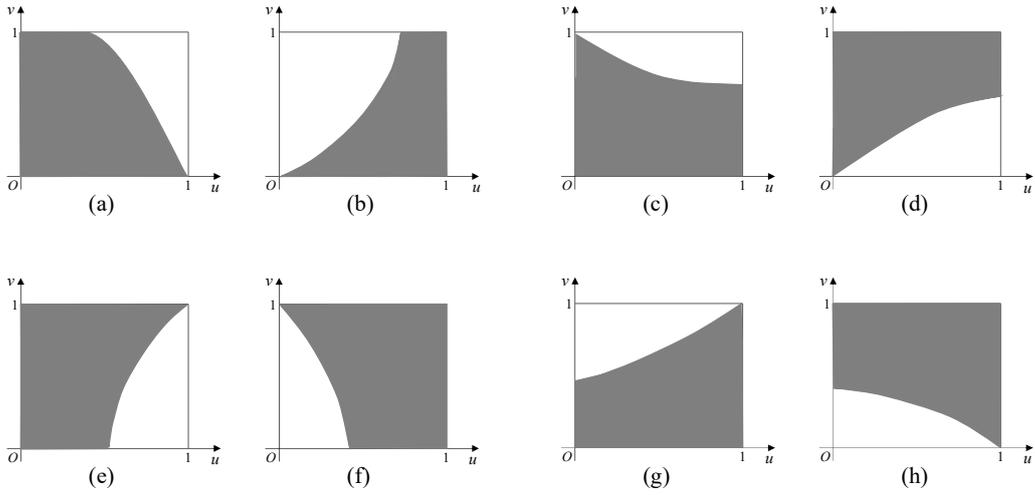

Figure 3. Eight cases of curved trapezoid domains of the boundary patch.

And the curved trapezoid domain limited by the boundary curve and the three isoparametric lines can be mapped into a standard parametric domain, we take Figure 3(a) as example, see Figure 4.

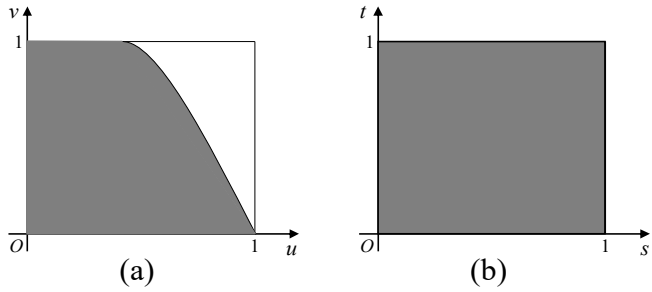

Figure 4. The mapping from curved trapezoid domain to the standard.

The reparameterization or normalization begins with a mapping, by which the irregular curved trapezoid domain shown in Figure 4(a) is mapped into the standard parametric domain shown in Figure 4(b), thus the surface on it can have a standard Bezier surface form with parametric domain [0,1]X[0,1] and therefore has a unified data structure.

Let the curve equation u= f(v) in Figure 4(a), the domain is u∈[0,f(v)], v∈[0,1], the following transformation maps [u,v]∈[0,f(v)]X[0,1] to [s,t]∈[0, 1]X[0, 1],

$$\Gamma: u = s * f(t), v = t \tag{2}$$

Where:

$$f(v) = a_p v^p + a_{p-1} v^{p-1} + \Lambda + a_0, p \in N, \forall a_i \in R, a_p \neq 0$$

is a polynomial of degree p.

Replace Eq. (2) into (1), we have

$$\begin{aligned} S(u,v) &= S((f(t) \cdot s), t) \\ &= \sum_{i=0}^{m}\sum_{j=0}^{n} B_{i,m}((a_p t^p + a_{p-1}t^{p-1} + \Lambda + a_0) \cdot s) B_{j,n}(t) R_{i,j} \\ &= \sum_{i=0}^{m}\sum_{j=0}^{m \times p+n} B_{i,m}(s) B_{j,m \times p+n}(t) \widetilde{R}_{i,j}; \end{aligned} \quad (3)$$

where, $\widetilde{R}_{i,j} \in \Re^3$ new control points.

In terms of parameter (s,t), the degree of the Bezier surface is from mXn up to mX(mXp+n). Depending on the accuracy requirements of different applications, the degree of polynomial f(v) may be degree 1, 2, or higher.

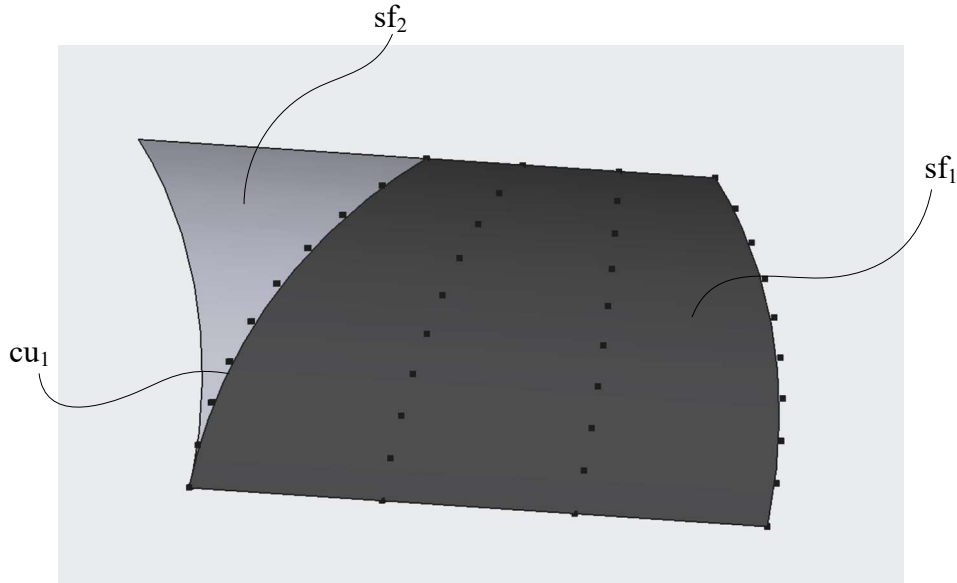

Figure 5. The reparameterized Bezier surface from 3X3 up to 3X9 with a quadric transformation, where the asterisk points are control points.

In Figure 5, shows a reparameterized or normalized surface, where sf1 is the part to be retained, sf2 is that to be cut-off, curve cu1 is the intersection or trimmed boundary, and the asterisk points are regenerated control points of new Bezier surface, which is defined on the standard parametric domain. Note that the shape of retained surface part and the boundary remain unchanged.

# 4. Replacement of boundary curves of trimmed surfaces by intersection curve

At this point, the two surfaces S1 and S2 have been transformed on the boundary and are expressed as a group of small Bezier surface patches, their shapes remain unchanged. Since the intersection curve C is not completely on the two trimmed surfaces. In general, the intersection curve C and the two on-surface curves SC1 and SC2 do not coincide with each other. There is either a gap or an overlap between two intersection surfaces, which is known as the "watertight" problem in academia and industry. [Sederberg et al., 2008], [Urick et al., 2019]

To solve this problem, we suggest to modify the boundary curves of reparameterized Bezier surface patches along the intersection border. The algorithm is simple, for the border control points of Bezier surface are the control points of border curves of the surface.

After the control points of the surface boundary curves at the intersections are replaced with control points of the intersection curve C, the two intersection surfaces have the curve C as the common boundary, and there are no gaps or overlaps anymore. As can be seen from Figure 6, the two intersection surfaces achieve a seamless connection, solving the watertight problem, thus making the product design closed and complete.

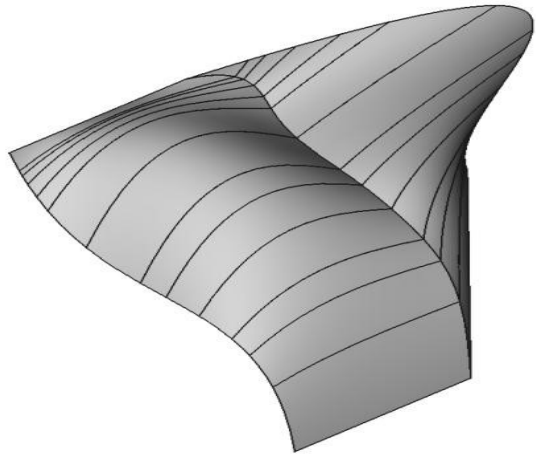

Figure 6. A watertighted example of two intersection Bezier surfaces, where the multi-parallel curves show the segmentation according to the intersection points.

It should be noted that the degree of the reparameterized surface will increase, the degree of the surface boundary curve on the intersection surface may be higher than the degree of the intersection curve C. The degree of the intersection curve C may be upgraded to the same degree as the reparameterized surface.

In practical applications, in order to improve the efficiency of design and calculation and to keep the same data structure, the degree of reparameterized surface patch may be appropriately reduced within the allowable range of error and accuracy.

## 5. Results, discussions, and conclusions

To sum up, there are two stages to realize the watertight:
1. reparameterizing, to transform the trapezoid domain to the standard one;
2. substituting, to replace the boundary curves of reparameterized surfaces by intersection curve.

In stage 1, the shape of trimmed surfaces keeps unchanged. In stage 2, the replacement will result in errors or bias at the boundary. To limit them in a range, higher order interpolation of intersection curve will need, so that more intersection points will be utilized to reach higher accuracy.

The technique is more suitable for the definition, representation and application of geometric models in fields such as CAD, CAE, CAM etc. Since the boundaries of surface of a product or object are standardized with a unified mathematical model and the same data structure, which facilitates parallel processing and quick visualization. The geometric models will be more easily converted and interpreted in different systems, supporting data sharing among suppliers, customers or collaborators, and reducing the risk of misunderstandings and errors. Furthermore, the method automatically processes intersection surface boundaries, generates a complete and closed model without gaps, reduces errors at the model surface boundaries, improves the accuracy at the model surface boundaries, and makes the product design generated based on the model more accurate. It is exquisite and speeds up the progress of design engineers in completing tasks. It not only facilitates subsequent analysis and processing and manufacturing processes, reduces risks in the product development process, but also shortens the development cycle and improves productivity.

Therefore, the solution can improve the quality and reliability of the product geometric model and is conducive to the smooth progress and efficient output of design, analysis and manufacturing processes, and has broad application prospects in CAD.

## 6.Acknowledgements

The authors would like to thank Ms. Yue Pan for her help in preparing some figures in the paper and Mr. Chengxin Huo for some discussions.

# 7.References

[1]. W. Böhm, G. Farin, and J. Kahmann. A survey of curve and surface methods in CAGD. Computer Aided Geometric Design (1) 1:1-60，1984

[2]. G.Farin. Curves and Surfaces for CAGD: A Practical Guide, 5 ed.,Academic Press, San Diego, CA, USA, 2002

[3]. J. Kasik, W. Buxton, and D.R. Ferguson. Ten CAD challenges. IEEE Computer Graphics and Applications 25(2): 81–92, 2005

[4]. Les A. Piegl. Ten challenges in computer-aided design. Computer-Aided Design (37) 4:461-470, 2005

[5]. T. W.Sederberg, G. T.Finnigan, X. Li, H. Lin, and H.Ipson. Watertight trimmed NURBS. ACM Transactions on Graphics 27(3), 79:1–79:8, 2008

[6]. J.Cottrell, T.Hughes, and Y.Bazilevs. Isogeometric Analysis: Toward integration of CAD and FEA, John Wiley & Sons, Ltd., Chichester, WestSussex, UK, 2009

[7]. T.J.R.Hughes,J.A.Cottrell, and Y.Bazilevs. Isogeometric analysis: CAD, finite elements, NURBS, exact geometry and mesh refinement. Computer Methods in Applied Mechanics and Engineering 194(39–41):4135–4195, 2005

[8]. B.Marussig, and T. J. R.Hughes. A review of trimming in isogeometric analysis: Challenges, data exchange and simulation aspects. Archives of Computational Methods in Engineering 18(5): 463–481, 2018

[9]. T.W. Sederberg, J. Zheng, A. Bakenov, and A. Nasri. T-splines and T-NURCCs. ACM SIGGRAPH :477–484, 2003

[10]. T.W. Sederberg, D.L. Cardon, G.T. Finnigan, N.S. North, J. Zheng, and T. Lyche. Tspline simplification and local refinement. ACM SIGGRAPH :276–283, 2004

[11]. B.Urick. Reconstruction of tensor product spline surfaces to integrate surface-surface intersection geometry and topology while maintaining intersurface continuity. Ph.D. thesis, The University of Texas at Austin, Austin, TX, 2016

[12]. B.Urick, B.Marussig, E. Cohen, R. H. Crawford, T. J.R. Hughes, and R.F. Riesenfeld. Watertight boolean operations A framework for creating CAD-compatible gap-free editable solid models. Computer-Aided Design 115:147–160, 2019

[13]. B.Urick, R. H. Crawford, T. J. R. Hugheset, E. Cohen, and R. F. Riesenfeld. Reconstruction of trimmed NURBS surfaces for gap-free intersections. Journal of Computing and Information Science in Engineering 20(5):1-16, 2020